\documentstyle[12pt,titlepage,epsf]{article}  
\newcommand{\nin}{\noindent}
\newcommand{\be}{\begin{equation}}
\newcommand{\ee}{\end{equation}}
\newcommand{\bea}{\begin{eqnarray}}
\newcommand{\eea}{\end{eqnarray}}

\newcommand{\nonu}{\nonumber\\}

\newcommand{\ol}{\overline}

\begin{document}

\hfill   NTUA-97/00

\begin{center}

{\Large Vacuum polarization in thermal QED\\
with an external magnetic field}

\vspace{1cm}

J.Alexandre\footnote{jalex@central.ntua.gr}

Department of Physics, National Technical University of
Athens,

Zografou Campus, 157 80 Athens, Greece

\vspace{15mm}

Abstract
\end{center}

The one-loop vacuum polarization tensor is computed in QED with an external, 
constant, homogeneous magnetic field at finite temperature. The Schwinger
proper-time formalism is used and the computations
are done in Euclidian space. The well-known results are
recovered when the temperature and/or the magnetic field are switched off
and the effect of the magnetic field on the Debye screening is discussed.

\section*{Introduction}

The question of dynamical chiral symetry breaking 
in thermal $QED$ with an external
magnetic field (magnetic catalysis) has been studied 
in the context of the electroweak 
transition \cite{electroweak} and also, with $QED_3$ (in 2+1 dimensions),
in the framework of effective 
descriptions of planar superconductors \cite{superconductors},\cite{afk2}.
Recent studies of the magnetic catalysis at zero temperature \cite{afk1}
showed that it is essential to take into account the momentum
dependence of the fermion self-energy since the dynamical mass
given by the constant self-energy
approximation proved to be too small, 
by several orders of magnitude 
in the case of $QED$. These studies have been made with the analysis of
the gap equation provided by the Schwinger-Dyson equation, where 
the photon propagator was truncated at the one-loop level. 
The polarization tensor in the presence of an external magnetic field
was used in its lowest Landau level approximation,
as was done in \cite{miransky}.
The study of the magnetic catalysis at finite temperature
taking into account the momentum dependence of the 
fermion self-energy has been done in $QED_3$ \cite{afk2} but not
in $QED$ for which only the constant self-energy approximation
has been done \cite{gusynin}, \cite{lee}. 
With a study including the momentum dependence,       
we can still expect the critical temperature for 
the magnetic catalysis to be of the order of
the dynamical mass found at zero temperature \cite{gusynin}, 
but where the latter is given by a momentum-dependent
analysis as was made in \cite{afk1}. 
As a first step in this direction, we compute here
the one-loop polarization tensor in finite temperature
$QED$ in the presence of a external, constant, homogeneous magnetic field.

The computation will be done in Euclidian space, 
using the proper-time formalism 
introduced by Schwinger \cite{schwinger} which 
takes into account the complete
interaction between the fermion and the external, classical  field.
The same computation has been done at zero temperature in 
the paper \cite{tsai}
which will be often cited in the present article and the 
generalization to any external constant field is done in \cite{schubert},
using the 'string-inspired' technique.
We note that the derivation of the Heisenberg-Euler lagrangian
has been done at finite temperature with the same formalism
\cite{dittrich}, as well as the generalization to any 
external constant electromagnetic field \cite{gies}.

Section 1 will introduce the notations and recall the 
characteristics of fermions in an external magnetic field.
Section 2 will be devoted to the computation of the 44-component
of the polarization tensor: this presentation is chosen
for the sake of clarity since the external environmement
strongly breaks the symmetry between the Lorentz indices such
that the computation is not straightforward. The 
technical details of the method will be explained and we will recover the
well-known results in the limit where the temperature and/or the
magnetic field go to zero.
The other components will be computed in section 3 where the
transversality of the polarization tensor will be checked.
The section 4 will give the strong field approximation of the 
44-component of the polarization tensor, consitent with the lowest 
Landau level approximation. Finally, the conlusion will show the Debye screening 
obtained through these computations.

\section{Fermions in a constant magnetic field}

To fix our notations we shortly review here the characteristics of
fermions in a external, constant, homogeneous magnetic field 
at zero temperature. 

The model we are going to consider is described by the Lagrangian density:

\be
{\cal L}=-\frac{1}{4}F_{\mu\nu}F^{\mu\nu}+i\ol\Psi D_\mu\gamma^\mu\Psi 
-m\ol\Psi\Psi,
\ee

\nin where $D_\mu = \partial_\mu+i e A_\mu + i e A^{ext}_\mu$, 
$A_\mu$ is the abelian quantum gauge field, $F_{\mu \nu}$ its
corresponding field strength, and $A^{ext}_\mu$ describes 
the external magnetic field. 
We recall the usual definition $e^2\equiv 4\pi\alpha$.

We will choose the symmetric gauge for the external field ($\vec B$ is 
in the direction 3)

\be\label{symgauge}
A^{ext}_0(x)=0,~~A^{ext}_1(x)=-\frac{B}{2}x_2,
~~A^{ext}_2(x)=+\frac{B}{2}x_1,~~A^{ext}_3(x)=0
\ee

\nin for which we know from the work by Schwinger 
\cite{schwinger} that the fermion propagator is given by:

\be
S(x,y)=e^{iex^\mu A^{ext}_\mu(y)}\tilde S(x-y),
\label{sphase}
\ee

\nin where the translational invariant propagator $\tilde S$ 
has the following Fourier transform in the proper-time formalism:

\bea\label{schwingerrep}
\tilde S(p)&=&\int_0^\infty ds e^{is\left(p_0^2-p_3^2-p_\bot^2
\frac{\tan(|eB|s)}{|eB|s}-m^2\right)}\nonu
&\times&
\left[(p^0\gamma^0-p^3\gamma^3+m)(1+\gamma^1\gamma^2\tan(|eB|s))
-p^\bot\gamma^\bot
(1+\tan^2(|eB|s))\right]
\eea

\nin where $p^\bot=(p^1,p^2)$ is the transverse momentum and the 
same notation holds for the gamma matrices. 

Let us now turn to the finite temperature case.
We will note the fermionic Matsubara modes $\hat\omega_l=(2l+1)\pi T$ 
and the bosonic ones $\omega_n=2n\pi T$. The translational invariant part
of the bare fermion propagator reads in Euclidian space 
($p_0\to i\hat\omega_l$) and with 
the substitution $s\to -is$:

\bea\label{freeTprop}
\tilde S_l(\vec p)&=&
-i\int_0^\infty ds e^{-s\left(\hat\omega_l^2+p_3^2+p_\bot^2
\frac{\tanh (|eB|s)}{|eB|s}+m^2\right)}\nonu
&\times&
\left[(-\hat\omega_l\gamma^4-p^3\gamma^3+m)(1-i\gamma^1\gamma^2\tanh(|eB|s))
-p^\bot\gamma^\bot(1-\tanh^2(|eB|s))\right]
\eea

\nin where the Euclidian gamma matrices satisfy the 
anticommutation relation
$\{\gamma^\mu,\gamma^\nu\}=-2\delta^{\mu\nu}$, with $\mu,\nu=1,2,3,4$
and $\vec p=(p_\bot,p_3)$.

Finally, the one-loop polarization tensor is 

\be\label{poltensor}
\Pi^{\mu\nu}_n(\vec k)=
-4\pi\alpha T\int\frac{d^3\vec p}{(2\pi)^3}
\sum_{l=-\infty}^\infty\mbox{tr}\left\{\gamma^\mu \tilde S_l(\vec p)
\gamma^\nu\tilde S_{l-n}(\vec p-\vec k)\right\}
+Q^{\mu\nu}(k)
\ee

\nin where $Q^{\mu\nu}$, usually called the 'contact term', 
cancels the ultraviolet divergences and
therefore does not depend on the temperature or on the
magnetic field since these give finite effects. 
The addition of this contact term is equivalent to the 
addition of the counterterm $(1-Z_3)F^{\mu\nu}F_{\mu\nu}/4$
in the original Lagrangian and
the usual ultraviolet divergences appear in the 
proper-time formalism as singularities in $s=0$, as will be seen 
in the next section. With this proper-time method, 
a cut-off $0<\varepsilon<s$ provides
a gauge invariant regularization which will be used in what 
follows. The limit $\varepsilon\to 0$ will be taken after
computing the contact term $Q^{\mu\nu}$.\\
We note that the $A^{ext}_\mu$-dependent phase of the fermion propagator
does not contribute to the polarization tensor since in coordinate space

\be
\exp\left\{ie\left(x^\mu A^{ext}_\mu(y)+
y^\mu A^{ext}_\mu(x)\right)\right\}=1
\ee

\nin with the specific choice of gauge (\ref{symgauge}).
If we choose another potential $A^{ext}_\mu$, 
it is shown in \cite{gies} that the
change of gauge is equivalent to the introduction of a 
chemical potential.

\section{44-component}

With the expression (\ref{freeTprop}) of the fermion propagator, we obtain
for the 44-component of the polarization tensor
after the integration over $\vec p$

\bea
&&\Pi^{44}_n(\vec k)=
\frac{-2\alpha T}{\sqrt\pi}|eB|\int
\frac{dsd\sigma}{\sqrt{s+\sigma}(\tanh(|eB|s)+\tanh(|eB|\sigma))}\\
&&\times\sum_{l=-\infty}^\infty
e^{-\frac{k_\bot^2}{|eB|}
\frac{\tanh(|eB|s)\tanh(|eB|\sigma)}{\tanh(|eB|s)+\tanh(|eB|\sigma)}
-[(s+\sigma)(\hat\omega_l^2+m^2)
+s\omega_n(\omega_n-2\hat\omega_l)+\frac{s\sigma}{s+\sigma}k_3^2]}\nonu
&&\times
\Bigg[k_\bot^2\frac{\tanh(|eB|s)\tanh(|eB|\sigma)}
{(\tanh(|eB|s)+\tanh(|eB|\sigma))^2}(1-\tanh(|eB|s))
(1-\tanh(|eB|\sigma))\nonu
&&~~~~~~~~-|eB|\frac{(1-\tanh(|eB|s))(1-\tanh(|eB|\sigma))}
{\tanh(|eB|s)+\tanh(|eB|\sigma)}\nonu
&&+\left(\hat\omega_l(\hat\omega_l-\omega_n)-m^2+
\frac{s\sigma}{(s+\sigma)^2}k_3^2-
\frac{1}{2(s+\sigma)}\right)
\left(1+\tanh(|eB|s)\tanh(|eB|\sigma)\right)\Bigg]+Q^{44}(k)\nonumber
\eea

\nin In finite temperature computations,
one usually first does the summation over Matsubara modes and then
the integration over momenta. In this formalism, what is important 
as we will see below is to do the summation over Matsubara modes
before the integration over the proper-time parameters, when
the cut-off is removed (i.e. $\varepsilon\to 0$).
As in \cite{tsai}, we make the change of variable  $s=u(1-v)/2$ and
$\sigma=u(1+v)/2$ to obtain

\bea\label{pol1}
&&\Pi^{44}_n(\vec k)=
\frac{-\alpha T}{\sqrt\pi}|eB|
\int_\varepsilon^\infty du\sqrt{u}\int_{-1}^1 dv\sum_{l=-\infty}^\infty
e^{-\frac{k_\bot^2}{|eB|}\frac{\cosh\overline u-\cosh\overline uv}
{2\sinh\overline u}-
u[\hat\omega_l^2+m^2
+(1-v)\omega_n(\omega_n/2-\hat\omega_l)+\frac{1-v^2}{4}k_3^2]}\nonu
&\times&\left[\left(\hat\omega_l(\hat\omega_l-\omega_n)+
\frac{1-v^2}{4}k_3^2-\frac{1}{2u}
-m^2\right)\coth\overline u
-\frac{|eB|}{\sinh^2\overline u}+
k_\bot^2\frac{\cosh\overline u-\cosh\overline uv}{2\sinh^3\overline u}
\right]+Q^{44}(k)\nonumber
\eea

\nin where $\overline u=|eB|u$. We make the integration by parts over $u$
(we note $\phi(u)$ the exponent)

\be
|eB|\int_\varepsilon^\infty du e^{-\phi(u)}
\frac{\sqrt u}{\sinh^2\overline u}\longrightarrow
\int_\varepsilon^\infty du
e^{-\phi(u)}\sqrt u\coth\overline u
\left(\frac{1}{2u}-\frac{d\phi(u)}{du}\right)
\ee

\nin where we disregard the surface term \cite{tsai}. 
We then obtain the final expression

\bea\label{pol44}
&&\Pi^{44}_n(\vec k)=\frac{-\alpha T}{\sqrt\pi}|eB|
\int_\varepsilon^\infty du\sqrt u\int_{-1}^1 dv\sum_{l=-\infty}^\infty
e^{-\frac{k_\bot^2}{|eB|}\frac{\cosh\overline u-\cosh\overline uv}
{2\sinh\overline u}
-u[m^2+W_l^2+\frac{1-v^2}{4}(\omega_n^2+k_3^2)]}\\
&&\times
\left[\frac{k_\bot^2}{2}\frac{\cosh\overline uv-
v\coth\overline u\sinh\overline uv}{\sinh\overline u}
-\coth\overline u\left(\frac{1}{u}-2W_l^2+v\omega_nW_l-
\frac{1-v^2}{2}k_3^2\right)\right]+Q^{44}(k)\nonumber
\eea

\nin where $W_l=\hat\omega_l-\frac{(1-v)}{2}\omega_n$.
We note for the purpose of consistency
that the integrand in (\ref{pol44}) is an even function 
of the parameter $v$ since $W_l(-v)=-W_{n-l-1}(v)$ and therefore
$\sum_le^{-uW_l^2}$ is even in $v$, which ensures the symmetry
between the proper-times $s$ and $\sigma$. Thus it is important to 
perform the summation over Matsubara modes before doing the 
integrations over the proper-time parameters. Another reason to 
do the summation over Matsubara modes first
is to avoid artificial divergences in the temperature-dependent
part of the polarization tensor (which should be finite), as will
be seen at the end of this section.

Let us now determine the contact term $Q^{44}(k)$. Since it does not 
depend on the temperature or the magnetic field, it will be
determined after taking the limit $T\to 0$ and 
$|eB|\to 0$ of (\ref{pol44}).
If we set $T=0$ in (\ref{pol44}), 
we recover the zero-temperature results given 
in \cite{tsai} since the substitutions
$W_l\to p_4$ and $T\sum_l\to (2\pi)^{-1}\int dp_4$ lead to 

\bea
\lim_{T\to 0}&&T\sum_{l=-\infty}^\infty e^{-uW_l^2}=
\frac{1}{2\sqrt{\pi u}}\nonu
\lim_{T\to 0}&&T\sum_{l=-\infty}^\infty \left(\frac{1}{u}-2W_l^2\right)
e^{-uW_l^2}=0\nonu
\lim_{T\to 0}&&T\sum_{l=-\infty}^\infty v\omega_nW_l e^{-uW_l^2}=0
\eea

\nin and therefore ($\omega_n\to k_4$)

\bea\label{piindep}
&&\lim_{T\to 0}\Pi_n^{44}(\vec k)=
\frac{-\alpha|eB|}{4\pi}
\int_\varepsilon^\infty du\int_{-1}^1 dv
e^{-\frac{k_\bot^2}{|eB|}\frac{\cosh\overline u-\cosh\overline uv}
{2\sinh\overline u}
-u[m^2+\frac{1-v^2}{4}(k_4^2+k_3^2)]}\nonu
&&\times \left[k_\bot^2
\frac{\cosh\overline uv-v\coth\overline u\sinh\overline uv}
{\sinh\overline u}+k_3^2(1-v^2)\coth\overline u\right]
+Q^{44}(k)
\eea

\nin However it is important to keep a non zero fermion mass
if we wish to recover this infrared limit:
we will see in the conclusion that if we set $m=0$,
$\Pi^{44}_0(0)$ reaches
a non-zero value in the limit $T\to 0$. 
Thus we
can commute the limit $T\to 0$ and the integration over the
proper-time $u$ to find a consistent zero temperature result
only if $m\ne 0$, at least as long as $|eB|>0$.
This condition is consistent since the magnetic catalysis
generates a dynamical mass when the temperature is lower than
the critical one, for any value of
the gauge coupling, which has been proven at least for
strong magnetic fields \cite{miransky}. Therefore if the fermion
should be massless, the use of the dynamically generated
mass $m_{dyn}$ instead of $m=0$ in 
(\ref{schwingerrep}), corresponding to a resummation of 
graphs for the fermion propagator, would lead us back to the massive case.
Thus we will consider $m\ne 0$ in what follows.

If we now wish to take the limit of zero magnetic field
of (\ref{piindep}), we take $\overline u\to 0$ and obtain

\be
\lim_{T\to 0,|eB|\to 0}\Pi_n^{44}(\vec k)=
\frac{-\alpha}{4\pi}
\int_\varepsilon^\infty \frac{du}{u}\int_{-1}^1 dv      
e^{-u[m^2+\frac{1-v^2}{4}k^2]} 
(1-v^2)\vec k^2+Q^{44}(k)
\ee

\nin where $k^2=k_4^2+\vec k^2$.
Then if we take the contact term

\be
Q^{44}(k)=\frac{\alpha}{4\pi}
\int_\varepsilon^\infty \frac{du}{u}\int_{-1}^1 dv
e^{-um^2}(1-v^2)\vec k^2
\ee

\nin we obtain finally when $\varepsilon\to 0$

\bea
\lim_{T\to 0,|eB|\to 0}\Pi_n^{44}(\vec k)&=&
\frac{-\alpha}{4\pi}
\int_0^\infty \frac{du}{u}\int_{-1}^1 dv(1-v^2)
\left(e^{-u[m^2+\frac{1-v^2}{4}k^2]}
-e^{-um^2}\right)\vec k^2\nonu
&=&\frac{\alpha}{4\pi}
\int_{-1}^1 dv(1-v^2)
\ln\left(1+\frac{1-v^2}{4m^2}k^2\right)
\vec k^2
\eea

\nin which is a result obtained by standard methods
\cite{zuber} with the Feynman parameter $z=(1+v)/2$.

To finish the comparison with results already established,
let us take the zero magnetic field 
limit of (\ref{pol44}). 
The limit $\overline u\to 0$ 
leads to

\bea\label{limebzero}
\lim_{|eB|\to 0}\Pi^{44}_n(\vec k)&=&\frac{-\alpha T}{\sqrt\pi}
\int_\varepsilon^\infty \frac{du}{\sqrt u}\int_{-1}^1 dv
\sum_{l=-\infty}^\infty
e^{-u[m^2+W_l^2+\frac{1-v^2}{4}(\omega_n^2+\vec k^2)]}\nonu
&\times&
\left[\frac{\vec k^2}{2}(1-v^2)
-\frac{1}{u}+2W_l^2-v\omega_nW_l\right]+Q^{44}(k)
\eea

\nin For $|eB|=0$, we can take a massless fermion ($m=0$) 
since there is no magnetic catalysis and
the Debye screening is then given by

\be\label{thermmB0}
M_{|eB|=0,m=0}^2(T)=-\lim_{\vec k^2\to 0}\Pi^{44}_0(\vec k)|_{|eB|=0}
=c~\alpha T^2
\ee

\nin with ($\varepsilon\to 0$)

\be\label{constant}
c=2\sqrt\pi\int_0^\infty \frac{du}{\sqrt u}
\sum_{l=-\infty}^\infty e^{-u(2l+1)^2}\left[2(2l+1)^2-\frac{1}{u}\right]
\ee 

\nin To compute $c$, we use the Poisson resummation 
\cite{dittrich}:

\be\label{poiss}
\sum_{l=-\infty}^\infty e^{-a(l-z)^2}=
\left(\frac{\pi}{a}\right)^{1/2}
\sum_{l=-\infty}^\infty e^{-\frac{\pi^2l^2}{a}-2i\pi zl}
\ee

\nin which shows that it is essential to perform the summation over
Matsubara modes before doing the integration over
the proper-time $u$ to avoid the
singularity $\int duu^{-3/2}$ in (\ref{constant}) since
we obtain 

\be
\sum_{l=-\infty}^\infty e^{-u(2l+1)^2}
\left[2(2l+1)^2-\frac{1}{u}\right]
=\frac{\pi^{5/2}}{2 u^{5/2}} 
\sum_{l\ge 1} (-1)^{l+1}l^2 e^{-\frac{\pi^2l^2}{4u}}
\ee

\nin such that

\be
c=\pi^3\int_0^\infty\frac{du}{u^3}\sum_{l\ge 1}(-1)^{l+1}
l^2e^{-\frac{l^2\pi^2}{4u}}
=\frac{16}{\pi}\int_0^\infty dx x e^{-x}
\sum_{l\ge 1}\frac{(-1)^{l+1}}{l^2}
=\frac{4\pi}{3}
\ee

\nin which gives the well known
result for the one-loop Debye screening with massless
fermions \cite{lebellac}
for which higher order corrections can be found in \cite{blaizot}.
We note that we can commute the integration over the proper-time
and the summation over the Matsubara modes after doing the Poisson 
resummation.

Using again the Poisson resumation (\ref{poiss}),
we can give another form of $\Pi_n^{44}(\vec k)$ which 
splits the temperature independent part from the temperature dependent one.
A straightforward computation leads to

\be\label{pitot}
\Pi_n^{44}(\vec k)=\Pi_n^0(\vec k)+\Pi_n^T(\vec k)
\ee

\nin where $\Pi_n^0(\vec k)$ is the zero temperature part (\ref{piindep})
(with $k_4\to\omega_n$) 
and $\Pi^T_n(\vec k)$ the temperature dependent part

\bea\label{pidep}
&&\Pi^T_n(\vec k)=\frac{-\alpha}{2\pi}|eB|
\int_0^\infty du\int_{-1}^1 dv
e^{-\frac{k_\bot^2}{|eB|}\frac{\cosh\overline u-\cosh\overline uv}
{2\sinh\overline u}
-u[m^2+\frac{1-v^2}{4}(\omega_n^2+k_3^2)]}\\
&&\times\sum_{l\ge 1}(-1)^le^{-\frac{l^2}{4uT^2}}
\left[\left(k_\bot^2\frac{\cosh\overline uv-
v\coth\overline u\sinh\overline uv}{\sinh\overline u}+
k_3^2(1-v^2)\coth\overline u\right)\cos\pi nl(1-v)\right.\nonu
&&\left.~~~~~~~~~~~~~~~~~~~~~~~~
-\frac{\coth\overline u}{u}\left(\frac{l^2}{uT^2}\cos\pi nl(1-v)
-2\pi vnl\sin\pi nl(1-v)\right)\right]\nonumber
\eea

\nin where we took $\varepsilon\to 0$ since the temperature dependent
part is finite. We see again that after this Poisson resumation every term 
of the Masubara series gives a finite integration over the 
proper-time $u$.

\section{Other components and transversality}

We now compute the other components in a similar way and 
will give only the important steps.
We first give the diagonal components of the polarization tensor
which all need integrations by parts to lead to the good limit 
when $T\to 0$. We set

\be
\phi_l(u,v)=\frac{k_\bot^2}{2|eB|}\frac{\cosh\overline u-
\cosh\overline uv}{\sinh\overline u}
+u\left[m^2+W_l^2+\frac{1-v^2}{4}(\omega_n^2+k_3^2)\right]
\ee

Let us start with $\Pi^{33}_n(\vec k)$:
the same steps as the ones used for the computation of
$\Pi^{44}_n(\vec k)$ and the same 
integration by parts lead to

\bea\label{pol33}
&&\Pi^{33}_n(\vec k)=\frac{-\alpha T}{\sqrt\pi}|eB|
\int_\varepsilon^\infty du\sqrt u\int_{-1}^1 dv
\sum_{l=-\infty}^\infty e^{-\phi_l(u,v)}\nonu
&&\times\left[v\omega_nW_l\coth\overline u
+\frac{k_\bot^2}{2}\frac{\cosh\overline uv-
v\coth\overline u\sinh\overline uv}{\sinh\overline u}
+\omega_n^2\frac{1-v^2}{2}\coth\overline u\right]+Q^{33}(k)
\eea

The computation of $\Pi_n^{ii}(\vec k)$, $i=1,2$ (without summation 
over $i$) is slightly different. 
After the integration over the loop momentum $\vec p$,
the change of variable $s=u(1-v)/2$
and $\sigma=u(1+v)/2$ leads to

\bea
\Pi^{ii}_n(\vec k)&=&\frac{-\alpha T}{\sqrt\pi}|eB|
\int_\varepsilon^\infty du\sqrt u\int_{-1}^1 dv
\sum_{l=-\infty}^\infty e^{-\phi_l(u,v)}
\left[\frac{\cosh\overline uv}{\sinh\overline u}
\left(v\omega_nW_l-\frac{1}{2u}
-W_l^2-m^2\right)\right.\nonu
&-&
\left.\frac{2k_i^2-k_\bot^2}{2}\frac{\cosh\overline u-
\cosh\overline uv}{\sinh^3\overline u}
+(\omega_n^2+k_3^2)\frac{1-v^2}{4}\frac{\cosh\overline uv}
{\sinh\overline u}\right]
+Q^{ii}(k)
\eea

\nin Then we make the integration by parts over $u$

\bea
&&\int_\varepsilon^\infty du
e^{-\phi_l(u,v)}m^2\sqrt u\frac{\cosh\overline uv}{\sinh\overline u}
\longrightarrow\\
&&\int_\varepsilon^\infty du
e^{-\phi_l(u,v)}\sqrt u\frac{\cosh\overline uv}{\sinh\overline u}
\left[\frac{1}{2u}+|eB|(v\tanh\overline uv-\coth\overline u)
-\frac{d}{du}\left(\phi_l(u,v)-
um^2\right)\right]\nonumber
\eea

\nin followed by the intergation by parts over $v$

\bea
&&\int_{-1}^1dve^{-\phi_l(u,v)}\frac{\cosh\overline uv}{\sinh\overline u}
\left[\frac{1}{u}
+|eB|\left(v\tanh\overline uv-
\coth\overline u\right)\right]
\longrightarrow\nonu
&&\int_{-1}^1dve^{-\phi_l(u,v)}\frac{v\cosh\overline uv
-\coth\overline u \sinh\overline uv}{\sinh\overline u}
\left[\omega_nW_l-\frac{k_\bot^2}{2}
\frac{\sinh\overline uv}{\sinh\overline u}-
\frac{v}{2}(\omega_n^2+k_3^2)\right]
\eea

\nin where we again disregarded the surface terms. We finally obtain 

\bea\label{polii}
&&\Pi^{ii}_n(\vec k)=\frac{-\alpha T}{\sqrt\pi}|eB|
\int_\varepsilon^\infty du\sqrt u\int_{-1}^1 dv
\sum_{l=-\infty}^\infty e^{-\phi_l(u,v)}
\left[\coth\overline u\frac{\sinh\overline uv}{\sinh\overline u}
\omega_nW_l\right.\nonu 
&+&
\left.(k_\bot^2-k_i^2)\frac{\cosh\overline u-\cosh\overline uv}
{\sinh^3\overline u}
+\frac{\omega_n^2+k_3^2}{2}\frac{\cosh\overline uv-
v\coth\overline u\sinh\overline uv}{\sinh\overline u}\right]
+Q^{ii}(k)
\eea

Now let us go to the off-diagonal components of the polarization tensor.
What differs from the diagonal components is that we do not make any 
integration by parts and we obtain directly the final results with the 
expected limit when $T\to 0$:

\bea\label{polother}
\Pi^{34}_n(\vec k)&=&\frac{\alpha T}{\sqrt\pi}|eB|
\int_\varepsilon^\infty du\sqrt u\int_{-1}^1 dv
\sum_{l=-\infty}^\infty e^{-\phi_l(u,v)}
k_3\left[vW_l+\frac{1-v^2}{2}\omega_n\right]\coth\overline u
+Q^{34}(k)\nonu
\Pi^{12}_n(\vec k)&=&\frac{\alpha T}{\sqrt\pi}|eB|
\int_\varepsilon^\infty du\sqrt u\int_{-1}^1 dv
\sum_{l=-\infty}^\infty e^{-\phi_l(u,v)}
k_1k_2\frac{\cosh\overline u-\cosh\overline uv}{\sinh^3\overline u}
+Q^{12}(k)\nonu
\Pi^{i4}_n(\vec k)&=&\frac{\alpha T}{\sqrt\pi}|eB|
\int_\varepsilon^\infty du\sqrt u\int_{-1}^1 dv
\sum_{l=-\infty}^\infty e^{-\phi_l(u,v)}\\
&&~~~~~~~~~~~~~~~~\times
k_i\left[W_l\coth\overline u\frac{\sinh\overline uv}
{\sinh\overline u}
+\frac{\omega_n}{2}\frac{\cosh\overline uv-
v\coth\overline u\sinh\overline uv}{\sinh\overline u}\right]
+Q^{i4}(k)\nonu
\Pi^{i3}_n(\vec k)&=&\frac{\alpha T}{\sqrt\pi}|eB|
\int_\varepsilon^\infty du\sqrt u\int_{-1}^1 dv
\sum_{l=-\infty}^\infty e^{-\phi_l(u,v)}
\frac{k_ik_3}{2}\frac{\cosh\overline uv-
v\coth\overline u\sinh\overline uv}{\sinh\overline u}
+Q^{i3}(k)\nonumber
\eea

\nin where $i=1,2$. It is easy to check that all the components of 
the polarization tensor
give the results found in \cite{tsai} when $T\to 0$. The contact terms 
are determined in the same way as $Q^{44}(k)$
and can be summarised as

\bea\label{ct}
Q^{\mu\nu}(k)&=&\frac{\alpha}{4\pi}
\int_\varepsilon^\infty \frac{du}{u}\int_{-1}^1 dv
e^{-um^2}(1-v^2)
\left(\delta^{\mu\nu}k^2-k^\mu k^\nu\right)\nonu
&=&\frac{\alpha}{3\pi}
\int_\varepsilon^\infty \frac{du}{u}
e^{-um^2}
\left(\delta^{\mu\nu}k^2-k^\mu k^\nu\right)
\eea

\nin as in \cite{tsai}.

It is important now to check the transversality of the polarization tensor,
which is not obvious since $\Pi^{\mu\nu}$ contains terms which are not
explicitely proportional to any external momentum component.
The contact term (\ref{ct}) is obviously transverse and
with the expressions (\ref{pol44}), (\ref{pol33}), (\ref{polii}) and
(\ref{polother}), we obtain 

\bea\label{transverse}
k_\nu\Pi_n^{\nu\mu}(\vec k)&=&
\omega_n\Pi^{4\mu}_n(\vec k)+k_3\Pi^{3\mu}_n(\vec k)+
k_i\Pi^{i\mu}_n(\vec k)\nonu
&=&\delta^{4\mu}\frac{\alpha T}{\sqrt\pi}|eB|
\int_\varepsilon^\infty du\sqrt u\coth\overline u\int_{-1}^1 dv
\sum_{l=-\infty}^\infty e^{-\phi_l(u,v)}\nonu
&&~~~~~~~~~~~~\times\left[\frac{\omega_n}{u}
+W_l\left(v(\omega_n^2+k_3^2)+
\frac{\sinh\overline uv}{\sinh\overline u}k_\bot^2-
2\omega_nW_l\right)\right]\nonu
&=&2\delta^{4\mu}\frac{\alpha T}{\sqrt\pi}|eB|
\int_\varepsilon^\infty \frac{du}{\sqrt u}\coth\overline u\int_{-1}^1 dv
\frac{d}{dv}\left(\sum_{l=-\infty}^\infty W_le^{-\phi_l(u,v)}\right)\nonu
&=&\mbox{surface term}
\eea

\nin so that the polarization tensor is transverse, since the above sum
is zero up to surface terms which are normally omitted in this formalism.

\section{Strong field approximation}

We give here the strong field approximation of the 44-component of
the polarization tensor that can be used in a strong field study
of the magnetic catalysis. 

The strong field asymptotic form of the fermion propagator (\ref{freeTprop}) can be 
found by taking the limit $|eB|\to\infty$ in the integrand, neglecting 
the shrinking region of integration where the product $s|eB|$ goes to 
zero \cite{calucci}. We obtain then

\bea\label{propLLL}
\tilde S_l(\vec p)&\simeq& -i\int_0^\infty ds e^{-(\hat\omega_l^2+p_3^2+
\frac{p_\bot^2}{|eB|s}+m^2)}
\left(-\hat\omega_l\gamma^4-p^3\gamma^3+m\right)\left(1-i\gamma^1\gamma^2\right)\nonu
&=&ie^{-\frac{p_\bot^2}{|eB|}}\frac{\hat\omega_l\gamma^4+p^3\gamma^3-m}
{\hat\omega_l^2+p_3^2+m^2}\left(1-i\gamma^1\gamma^2\right)
\eea

\nin which is the well-known lowest Landau level approximation for the fermion 
propagator \cite{miransky} that we can obtain by truncating the expansion of
the propagator over the Landau levels to the dominant term \cite{chodos}. 
We will take the same limit $|eB|\to\infty$ in the expressions (\ref{piindep}) and 
(\ref{pidep}) to find the asymptotic form of $\Pi^{44}$. We note that the polarization
tensor does not contain divergences in the limit $|eB|\to\infty$, the magnetic field 
playing the role of a cut-off for the momenta. This is due to the exponential decrease
of the transverse degrees of freedom as can be seen in (\ref{propLLL}). Therefore we will not
consider any contact term here.

The temperature-independent part (\ref{piindep}) reads in the strong field limit 
($\overline u\to \infty$) 

\bea
\Pi_n^0(\vec k)&\simeq& -\frac{\alpha |eB|}{4\pi}\int_0^\infty du\int_{-1}^1 dv
e^{-\frac{k_\bot^2}{2|eB|}-u[m^2+\frac{1-v^2}{4}(\omega_n^2+k_3^2)]}\nonu
&&\times\left\{k_\bot^2\left[(1-v)e^{-\overline u(1-v)}+(1+v)e^{-\overline u(1+v)}\right]
+k_3^2(1-v^2)\right\}\nonu
&=&-\frac{\alpha |eB|}{4\pi}e^{-\frac{k_\bot^2}{2|eB|}}\int_{-1}^1 dv\left[
\frac{(1-v^2)k_3^2}{m^2+\frac{1-v^2}{4}(\omega_n^2+k_3^2)}+
{\cal O}\left(\frac{k_\bot^2}{|eB|}\right)\right]
\eea

\nin The integration over $v$ of the dominant term leads then 
to the following expression that was already 
derived in \cite{calucci} where the authors started the computation 
with the propagator (\ref{propLLL}):

\be
\Pi_n^0(\vec k)\simeq-\frac{2\alpha}{\pi}|eB|\frac{k_3^2}{k_\|^2}
e^{-\frac{k_\bot^2}{2|eB|}}
\left[1-\frac{2m^2}{\sqrt{k_\|^2(4m^2+k_\|^2)}}
\ln\left(\frac{\sqrt{4m^2+k_\|^2}+\sqrt{k_\|^2}}
{\sqrt{4m^2+k_\|^2}-\sqrt{k_\|^2}}\right)\right]
\ee

\nin where $k_\|^2=\omega_n^2+k_3^2$.

For the temperature-dependent part (\ref{pidep}), the limit $\overline u\to \infty$ 
followed by the change of variable $u\to u/|eB|$ leads to the dominant term

\be
\Pi_n^T(\vec k)\simeq-\frac{\alpha}{2\pi}\frac{|eB|^2}{T^2}e^{-\frac{k_\bot^2}{2|eB|}}
\int_0^\infty \frac{du}{u^2}\int_{-1}^1 dv 
e^{-u\frac{\mu^2(v)}{4|eB|}}
\sum_{l\ge 1}(-1)^{l+1}l^2 e^{-\frac{l^2|eB|}{4uT^2}}
\cos\pi nl(1-v)
\ee

\nin where $\mu^2(v)=4m^2+(1-v^2)k_\|^2$. We recognize here the Bessel function $K_1$ since

\be
\int_0^\infty \frac{du}{u^2} e^{-au-\frac{b}{u}}=
\sqrt\frac{a}{b}\int_0^\infty due^{-(u+\frac{1}{u})\sqrt{ab}}
=2\sqrt\frac{a}{b}K_1(2\sqrt{ab})
\ee

\nin where we defined

\be
a=\frac{\mu^2(v)}{4|eB|}~~~~\mbox{and}~~~~b=\frac{l^2|eB|}{4T^2}
\ee

\nin The temperature-dependent part can finaly be written

\be
\Pi_n^T(\vec k)\simeq-\frac{2\alpha}{\pi}|eB|e^{-\frac{k_\bot^2}{2|eB|}}
\sum_{l\ge 1}(-1)^{l+1}\int_{-1}^1 dv
\frac{l\mu(v)}{2T}K_1\left(\frac{l\mu(v)}{2T}\right)\cos\pi nl(1-v)
\ee

\nin We will see in the conclusion that the previous sum and integrals can be 
evaluated for the computation of the Debye screening in the regime where 
$m<<T<<\sqrt{|eB|}$.

\section*{Conclusion: Debye screening in a magnetic field}

To conclude, we look in more details at the Debye screening obtained
in this computation. From equation (\ref{pidep}), 
we find for the Debye mass

\be\label{debyem}
M_{|eB|}^2(T)=-\lim_{\vec k^2\to 0}\Pi^{T}_0(\vec k)
=\frac{\alpha|eB|}{\pi T^2}\int_0^\infty \frac{du}{u^2}\coth\overline u
e^{-um^2}\sum_{l\ge 1}(-1)^{l+1}l^2e^{-\frac{l^2}{4uT^2}}
\ee

\nin The zero-magnetic field limit is

\be
M_{|eB|=0}^2(T)=
\frac{\alpha}{\pi T^2}\int_0^\infty \frac{du}{u^3}
e^{-um^2}\sum_{l\ge 1}(-1)^{l+1}l^2e^{-\frac{l^2}{4uT^2}}
\ee

\begin{figure}
\epsfxsize=10cm
\epsfysize=8cm
\centerline{\epsfbox{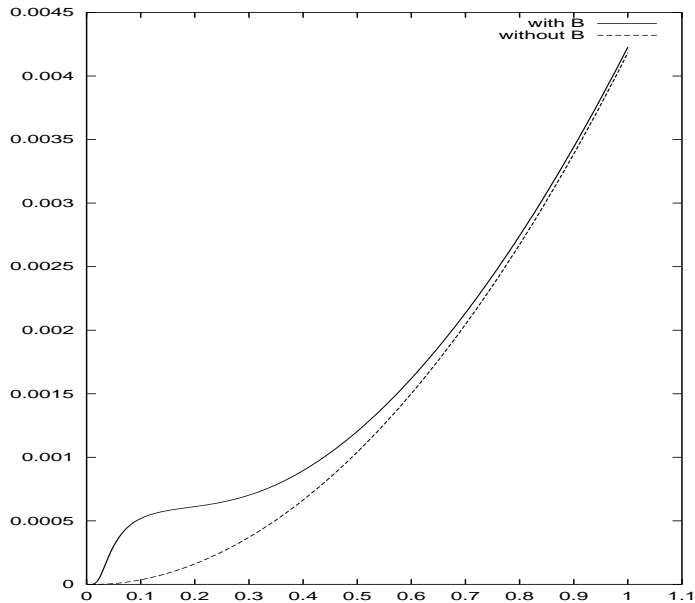}}
\caption{$M_{|eB|}^2/|eB|$ and $M_{|eB|=0}^2/|eB|$ versus 
$T/\sqrt{|eB|}$ for $\alpha=.001$ and $m/\sqrt{|eB|}=.1$}
\label{thma}
\end{figure}

\nin For given values of $|eB|$ and $m$, 
in figure \ref{thma} we compare
the ratios $M_{|eB|}^2/|eB|$ and $M_{|eB|=0}^2/|eB|$ as functions
of $T/\sqrt{|eB|}$, such that all the dimensionful quantities
are rescaled in units of the magnetic field ($[eB]=2$).
For high temperatures the curves converge towards the  
result (\ref{thermmB0}) (rescaled by $|eB|$)
since $m<<T$ and $\sqrt{|eB|}<<T$,
but for strong magnetic field
$T<<\sqrt{|eB|}$, a strong Debye sceening is generated compared to the
one without external field. As long as the temperature remains 
greater than the fermion mass, the Debye screening follows a plateau
when the temperature decreases.
We note that if we had $m=0$, the limit of the Debye screening
when $T\to 0$ would be (after the change of variable $u\to u/T^2$ in
(\ref{debyem}))

\bea\label{plateau}
\lim_{T\to 0}M_{|eB|,m=0}^2(T)&=&
\frac{\alpha|eB|}{\pi}\int_0^\infty \frac{du}{u^2}
\sum_{l\ge 1}(-1)^{l+1}l^2e^{-\frac{l^2}{4u}}\nonu
&=&\frac{4\alpha}{\pi}|eB|\int_0^\infty dx
\sum_{l\ge 1}(-1)^{l+1}le^{-xl}\nonu
&=&\frac{4\alpha}{\pi}|eB|\lim_{x\to 0}
\sum_{l\ge 1}(-1)^{l+1}e^{-xl}\nonu 
&=&\frac{4\alpha}{\pi}|eB|\lim_{x\to 0} 
\left(1-\frac{1}{1+e^{-x}}\right)\nonu
&=&\frac{2\alpha}{\pi}|eB|
\eea

\nin But when $T<m$ the fermion mass forces the screening
to vanish with the temperature, so that the value (\ref{plateau})
of the plateau is valid only if $m<<T<<\sqrt{|eB|}$. 

We note that a more unexpected behaviour has been observed in $QED_3$
at finite temperature 
in an external magnetic field \cite{afk2}: $M^2_{|eB|}$
first increases when the temperature decreases 
(in the region $T<<\sqrt{|eB|}$),
reaches a maximum when $T\simeq m$ and then decreases to 0 when $T\to 0$.

These behaviours of the Debye screening are a consequence of the dimensional 
reduction of the fermion dynamics in a strong magnetic field, as can be seen
with the propagator (\ref{propLLL}).

To conclude, we note again the consistency between the necessity to 
have a massive fermion to obtain the good zero temperature limits
(as long as $|eB|>0$) and the occurence of 
the magnetic catalysis which generates dynamically this mass.

\section*{Acknowledgements}

This work has been done within the
TMR project 'Finite temperature phase transitions in particle Physics',
EU contract number: FMRX-CT97-0122, and I would like to thank 
K.Farakos and G.Koutsoumbas for guiding me through the subject of
magnetic catalysis. Finally, I would like to thank G.Tiktopoulos 
for helpful advice.

\end{document}